\documentclass[superscriptaddress,showpacs,showkeys,preprintnumbers,nofootinbib,amsmath,amssymb,aps,onecolumn,prd,floatfix,11pt]{revtex4-2}
\usepackage{graphicx}

\begin{document}

\title{Primordial Black Holes in Induced Gravity Inflationary Models\\ with Two Scalar Fields}

\author{E.O.~Pozdeeva}\email{pozdeeva@www-hep.sinp.msu.ru}
\affiliation{Skobeltsyn Institute of Nuclear Physics, Lomonosov Moscow State University, Leninskie Gory 1, Moscow 119991,  Russia}
\author{S.Yu.~Vernov}\email{svernov@theory.sinp.msu.ru}
\affiliation{Skobeltsyn Institute of Nuclear Physics, Lomonosov Moscow State University, Leninskie Gory 1, Moscow 119991,  Russia}

\begin{abstract}
We propose a two-field inflationary model with the induced gravity term. Using conformal transformation of the metric,  we get the chiral cosmological model with two scalar fields. We demonstrate that the constructed inflationary model does not contradict to the recent observation data and is suitable for  primordial black hole (PBH) formation. The estimation of PBH masses allows to consider PBHs as dark matter candidates.
\end{abstract}


%PACS: 98.80.Cq; 04.50.Kd

% 04.50.Kd 	Modified theories of gravity
% 98.80.Cq Particle-theory and field-theory models of the early Universe (including cosmic pancakes, cosmic strings, chaotic phenomena, inflationary universe, etc.)


\maketitle

\label{sec:intro}
\section{Introduction}

The possibility that a significant fraction or even the totality of the dark matter is not a new form of matter but consists of primordial black holes (PBHs)
is actively discussed~~\cite{Dolgov:1992pu,Ivanov:1994pa,Garcia-Bellido:1996mdl,Khlopov:2008qy,Kamenshchik:2018sig,Green:2020jor,Carr:2020xqk}. The hypothesis of the existence of PBH is supported to the increasing amount of direct and indirect observations of black holes with masses beyond the astrophysical range, the occurrence of which is not described by models of stellar collapse. In particular, the large masses of the discovered black holes in the centres of galaxies suggest the hypothesis about the galaxies formations around previously existed black holes~\cite{Dolgov:2017aec}.

A black hole is called primordial if it was formed before the matter dominance epoch. The most popular PBH formation mechanism assumes the existence of the over-densities larger than some critical value arising during inflation. These over-densities may form PBHs during the radiation dominated era~\cite{Ozsoy:2023ryl}.
  Such scenarios are not realized in the slow-roll inflation, so the slow-roll conditions should be violated in the models that unify inflation and PBH formation. In inflationary models with one minimally coupled scalar field, two slow-roll parameter should be smaller than one in the slow-roll regime. When the first parameter becomes equal to one, inflation as an accelerated expansion of the Universe stops, so only the second slow-roll parameter can be more than one during inflation. In single-field inflationary models, PBH formation corresponds to an ultra slow-roll stage of inflation~\cite{Chataignier:2023ago,Firouzjahi:2023ahg}.

In many two-field models, one scalar field plays a role of inflaton in the beginning of inflation and another field plays the same role at the end. The investigations of such inflationary models with two stages of inflation show that density perturbations at the time corresponding to the transition between two inflationary stages can be so large that leads to PBH production~\cite{Garcia-Bellido:1996mdl,Pi:2017gih,Gundhi:2020kzm,Braglia:2020eai,Ketov:2021fww,Braglia:2022phb,Cheong:2022gfc,Kawai:2022emp}.

Induced gravity models are a natural generalization of General Relativity, in which quantum effects leading to a non-trivial coupling between a scalar field and the Ricci scalar are taken into
account~\cite{Elizalde:1993ew,Elizalde:2015nya,Tronconi:2017wps,Kamenshchik:2019alc}. In context of PBH production, single-field induced gravity models are studied in~\cite{Chataignier:2023ago,Kamenshchik:2024kay}.

In this paper, we propose a two-field inflationary model with the induced gravity term. Using conformal transformation of the metric, we get a chiral cosmological model~~\cite{Chervon:1995jx,Karananas:2016kyt,Chervon:2019nwq,Fomin:2020caa,Geller:2022nkr,Pozdeeva:2024lah} (CCM) with two scalar fields. We analyze the behaviour of scalar fields during inflation by numerical calculations and demonstrate that the constructed inflationary model do not contradict to observation data and is suitable for PBH formation. The estimation of PBH masses shows that PBHs can be considered as dark matter candidates.

\section{Induced gravity models and the corresponding CCMs}

Let us consider the induced gravity model with two scalar fields, described by the following action:
 \begin{equation}
S=\int d^4x\sqrt{-\tilde{g}}\left[\frac{\xi}{2}\sigma^2\tilde{R}-\frac12\tilde{g}^{\mu\nu}\partial_{\mu}\sigma\partial_{\nu}\sigma-\frac12\tilde{g}^{\mu\nu}\partial_{\mu}\chi\partial_{\nu}\chi-\tilde{V}(\sigma,\chi) \right],
\label{FR}
\end{equation}
where $\xi$ is a positive constant,  the potential $\tilde{V}$ is a differentiable function.

Using the conformal transformation of the metric
\begin{equation*}
g_{\mu\nu}=\frac{\xi\sigma^2}{M_\mathrm{Pl}^2}\tilde{g}_{\mu\nu},
\end{equation*}
where $M_\mathrm{Pl}$ is the Planck mass, we obtain the action of the two-field CCM:
\begin{equation}
\label{FRSE}
S_{E}=\int d^4x\sqrt{-g}\left[\frac{M_\mathrm{Pl}^2}{2}R-\frac{1}{2}g^{\mu\nu}\partial_\mu\phi\partial_\nu\phi-\frac{y}{2}{g^{\mu\nu}}\partial_\mu\chi\partial_\nu\chi-V_E\right],
\end{equation}
where
\begin{equation}
\label{phisigma}
      \phi=\,M_\mathrm{Pl}\sqrt{6+\frac{1}{\xi}\,}\,\ln\left(\frac{\sigma}{M_\mathrm{Pl}}\right),
\end{equation}
\begin{equation}
y=\frac{M_\mathrm{Pl}^2}{\xi\sigma^2}=\frac{1}{\xi}\exp\left(-2\sqrt{\frac{\xi}{6\xi+1}\,}\frac{\phi}{M_\mathrm{Pl}}\right),
\end{equation}
and the potential
\begin{equation}
\label{V_E}
V_E= y^2(\phi)\tilde{V}(\sigma(\phi),\chi).
\end{equation}

\section{Evolution equations in the Einstein frame}

In the spatially flat Friedmann--Lema\^{i}tre--Robert\-son--Walker metric with the interval
\begin{equation*}
{ds}^{2}={}-{dt}^2+a^2(t)\left(dx_1^2+dx_2^2+dx_3^2\right),
\end{equation*}
the evolution equations have the following form~\cite{Chervon:2019nwq}:
\begin{equation}
	\label{H2}
H^2=\frac{1}{6M_\mathrm{Pl}^2}\left(X^2+2V_E\right)\,,\qquad
\dot H={}-\frac{X^2}{2M_\mathrm{Pl}^2} \,,
\end{equation}
where dots denote the time derivatives, $X\equiv\sqrt{{\dot{\phi}}^2+y\,{\dot{\xi}}^2}$ and the Hubble parameter $H(t)$ is the logarithmic derivative of the scale factor: $ H=\dot a/{a}$.

The field equations are as follows:
\begin{equation}
\label{equphi}
\ddot{\phi}+3H\dot{\phi}-\frac12\, \frac{dy}{d\phi}\,\dot{\xi}^2+\frac{\partial V_E}{\partial \phi}=0\,,
\end{equation}
\begin{equation}
\label{equxi}
\ddot{\chi}+3H\dot{\chi}+\frac{1}{y}\, \frac{dy}{d\phi}\,\dot{\chi}\dot{\phi}+\frac{1}{y}\,\frac{\partial V_E}{\partial \chi}=0\,.
\end{equation}

It is suitable to consider the e-folding number
$N=\ln(a/a_{e})$, where $a_{e}$ is a constant, as an independent variable during inflation.
Using the relation $\frac{d}{dt}=H\,\frac{d}{dN}$, equations~(\ref{H2}) can be written as follows
\begin{equation}
	\label{a2N} H^2=\frac{2V_E}{6M_\mathrm{Pl}^2-{\phi^\prime}^2-y{\chi^\prime}^2}\,,
\end{equation}
\begin{equation}\label{dHdN}
 H'={}-\frac{H}{2M_\mathrm{Pl}^2}\left[{\phi^\prime}^2+y{\chi^\prime}^2\right]\,,
\end{equation}
where primes denote derivatives with respect to $N$.

The standard slow-roll parameters $\epsilon_1$ and $\epsilon_2$ are
\begin{eqnarray}
  \epsilon_1 &=&{}-\frac{H^\prime}{H}=\frac{1}{2M_\mathrm{Pl}^2}\left[{\phi^\prime}^2+y{\chi^\prime}^2\right], \\
  \epsilon_2 &=&\frac{\epsilon_1^\prime}{\epsilon_1}=2\epsilon_1+\frac{1}{HX^2}\,\frac{dX}{dt}.
\end{eqnarray}

Using Eqs.~(\ref{a2N}) and (\ref{dHdN}), we eliminate $H^2$ and $H'$ from the field equations and obtain the following system:
\begin{equation}
\label{DYNSYSTEMN}
\begin{split}
    \phi''=&(\epsilon_1-3)\phi'+\frac12\,\frac{dy}{d\phi}\,{\chi'}^2-\frac{6M_\mathrm{Pl}^{2}-y{\chi'}^2-{\phi'}^2}{2V_E}\,\frac{\partial V_E}{\partial \phi}\,,\\
    {\chi}''=&(\epsilon_1-3)\chi'-\frac{1}{y}\, \frac{dy}{d\phi}\,\chi'\phi'-\frac{6M_\mathrm{Pl}^2-{\phi'}^2-y{\chi'}^2}{2yV_E}\frac{\partial V_E}{\partial \chi}\,.\\
\end{split}
\end{equation}

\section{Inflationary model}
We consider the following potential:
\begin{equation}
\label{Vsigmachi}
    \tilde{V}(\sigma,\chi)=\lambda\sigma^4\left(F_1(\chi)+F_2(\chi)\mathrm{e}^{\gamma\,\left[\ln(\sigma/M_\mathrm{Pl})\right]^{2\alpha}}\right),
\end{equation}
where
\begin{equation}\label{FF}
    F_1(\chi)=\left(1-\frac{\chi^2}{\chi_0^2}\right)^2-d\frac{\chi}{\chi_0},\qquad F_2(\chi)=\frac{c_2\chi^2}{\chi_0^2}+c_0\,,
\end{equation}
$\alpha$, $\gamma$, $\lambda$, $\chi_0$, $c_0$, $c_2$ and $d$ are constants. Note that the potential $\tilde{V}$ is real even if $\ln(\sigma/M_\mathrm{Pl})<0$ and $\alpha$ is not an integer number.

In the Einstein frame, we get
\begin{equation}\label{VE}
   V_E(\phi,\chi)=V_0\left(F_1(\chi)+F_2(\chi)
    {\rm e}^{\beta\left(\frac{\phi^2}{M_\mathrm{Pl}^2}\right)^\alpha}\right),
\end{equation}
where
\begin{equation}\label{gam}
  V_0=\frac{\lambda M_\mathrm{Pl}^4}{\xi^2},\qquad \beta=\gamma\left({\frac{\xi}{1+6\,\xi}}\right)^{\alpha}\,.
\end{equation}

The slow-roll parameters $\epsilon_1$ and $\epsilon_2$ do not depend on $V_0$. So,
the inflationary parameters: the spectral index and the tensor-to-scalar ratio
\begin{equation}
\label{n_s_r} n_s=1-2\epsilon_1-\epsilon_2, \qquad
r\approx16\epsilon_1,
\end{equation}
do not depend on $V_0$ as well. This model parameter can be chosen to get the observation value of the amplitude of scalar perturbations
\begin{equation}
\label{PR}
A_s=\frac{2H^2}{\pi^2M_\mathrm{Pl}^2r}.
\end{equation}

\section{Numerical solutions of the evolution equations}

We solve system~(\ref{DYNSYSTEMN}) numerically to analyze the evolution of scalar fields during inflation and to get values of inflationary parameters $n_s$, $r$, and $A_s$.
We define the e-folding number $N$ in such a way that $N=0$ corresponds to the moment at which inflationary parameters are calculated.

Note that $\phi=0$ corresponds to $\sigma=M_\mathrm{Pl}$, so at $\xi=1$ the initial induced gravity model becomes close to general relativity at the second stage of inflation. By this reason, we fix $\xi=1$.
It follows from formula (\ref{gam}) that the condition  $\xi=1$ does not restrict the choice of the potential $ V_E(\phi,\chi)$. The parameter $V_0$ is fixed by the condition $A_s=2.10\times 10^{-9}$ at $N=0$.

\begin{figure}[htb]
 \includegraphics[width=0.27\linewidth]{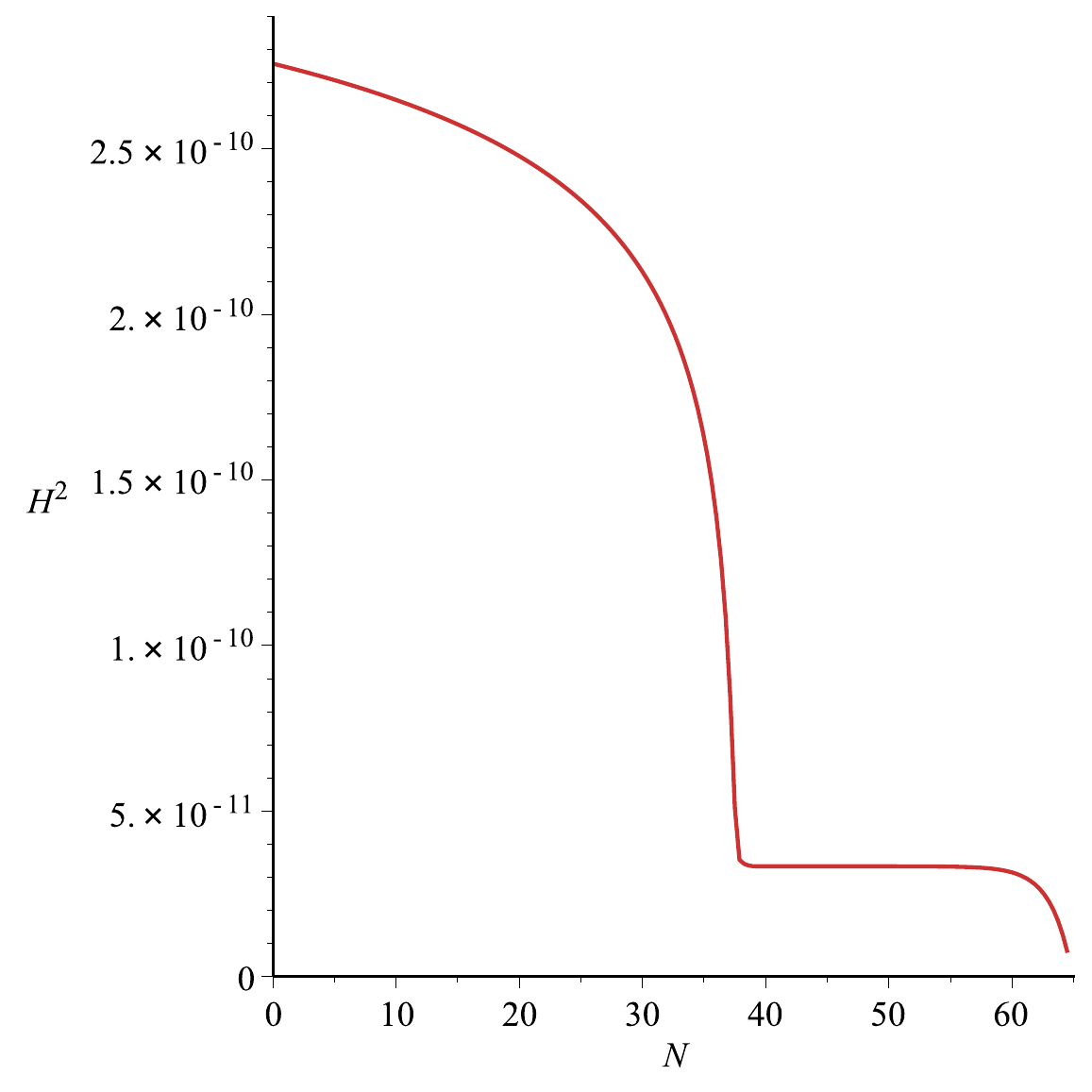} \
 \includegraphics[width=0.27\linewidth]{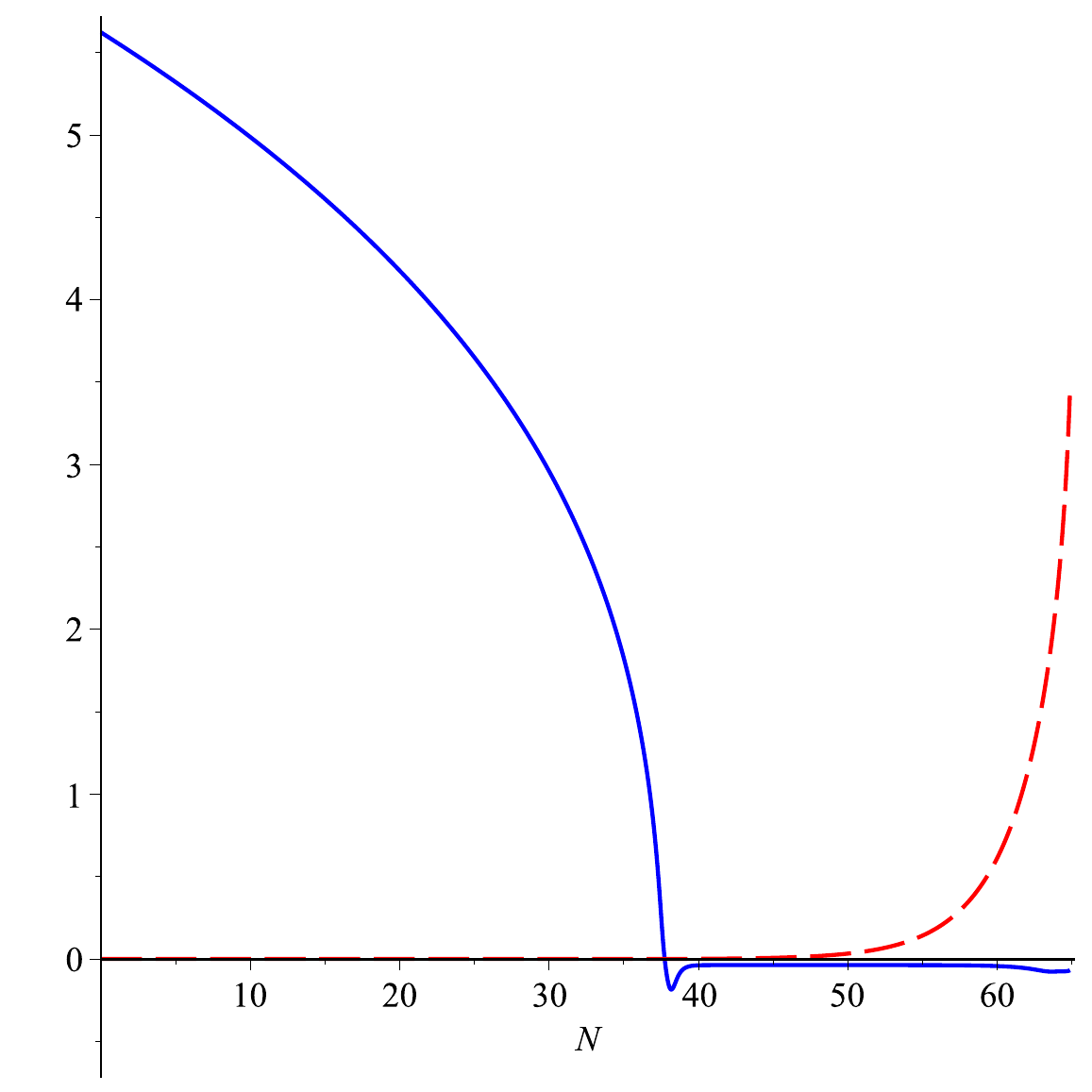} \
 \includegraphics[width=0.43\linewidth]{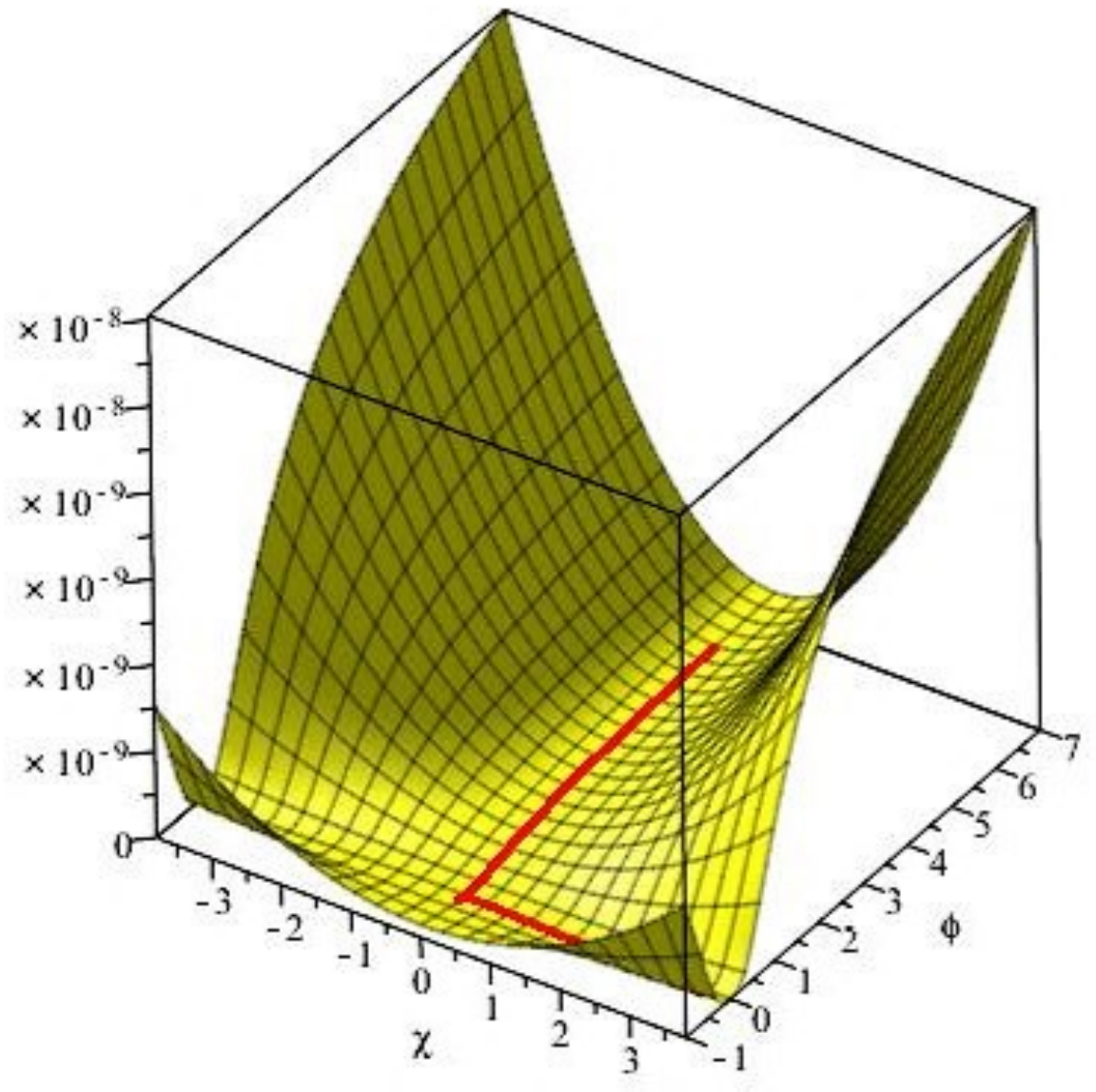}
\caption{The behaviour of $H^2(N)$ (left picture) and scalar fields $\phi(N)$ (blue solid curve) and $\xi(N)$ (red dash curve) during inflation (centre picture). Values of the model parameters are given in~\eqref{modelparam}.
The potential $V(\phi,\xi)$ and the trajectory are shown on the right picture. The Hubble parameter and the fields are shown in units~of~$M_\mathrm{Pl}$.
\label{Fig1Hfields}}
\end{figure}

\begin{figure}[htb]
 \includegraphics[width=0.31\linewidth]{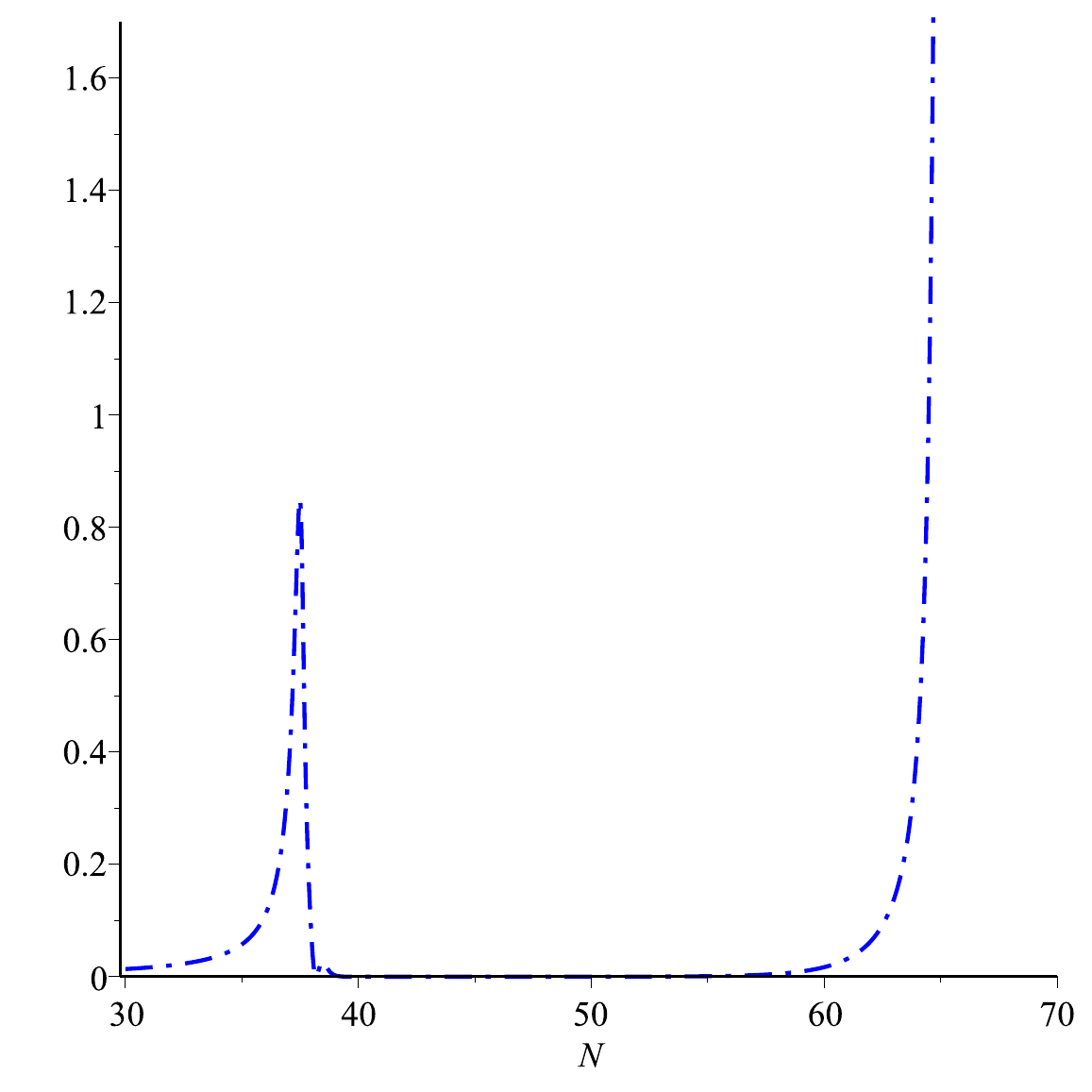} \
 \includegraphics[width=0.31\linewidth]{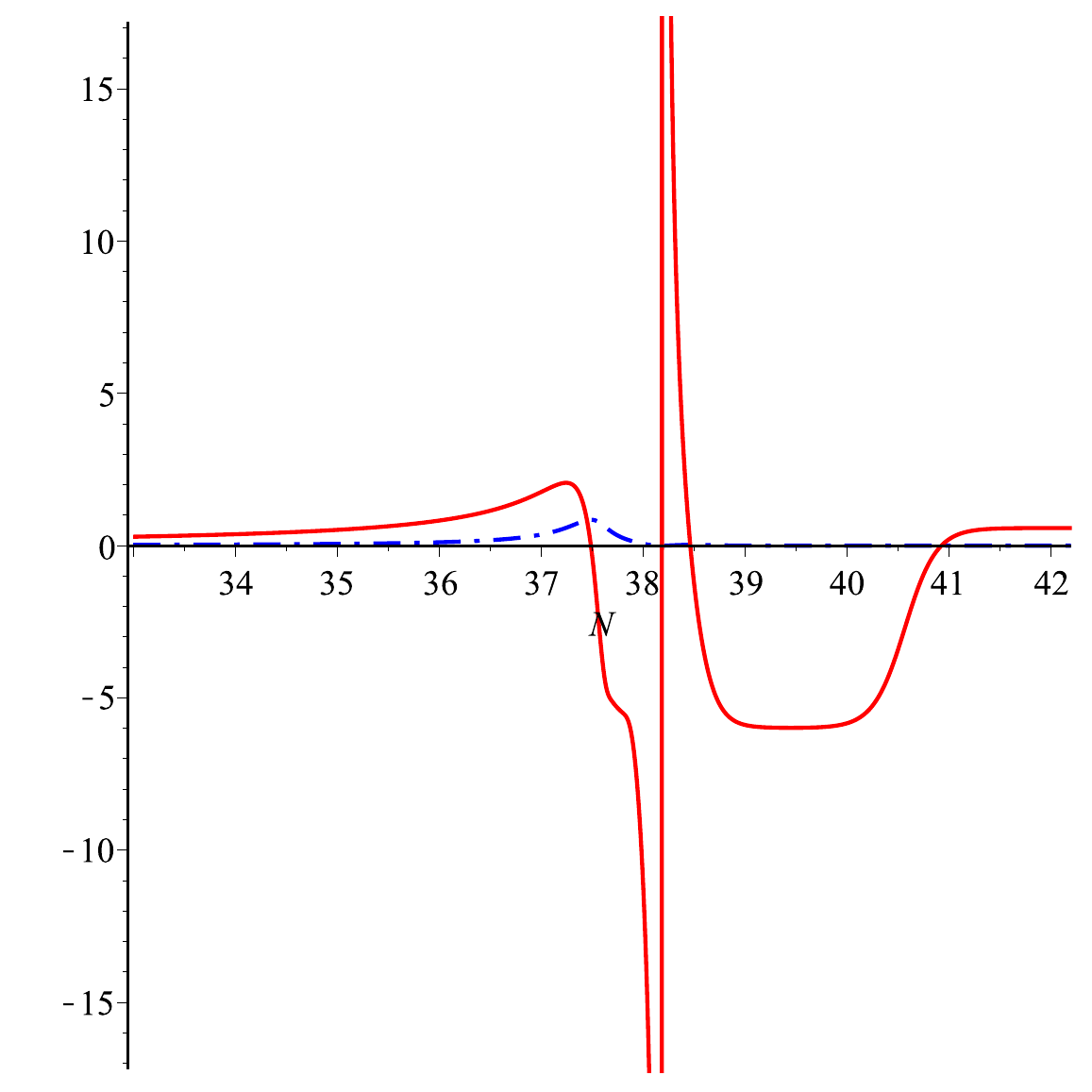} \
  \includegraphics[width=0.31\linewidth]{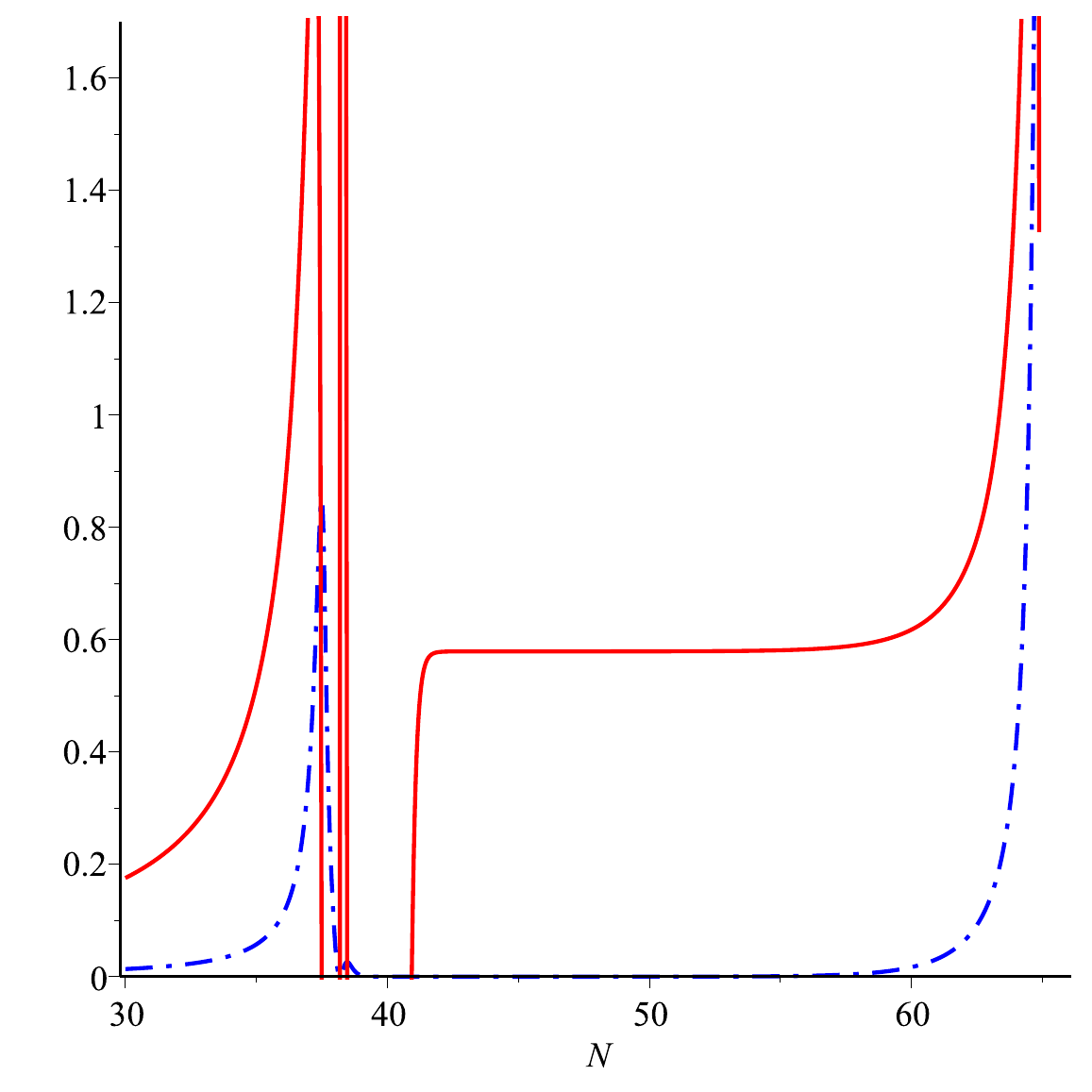}
\caption{The behavior of slow-roll parameters $\epsilon_1$ (blue dash-dot curve) and $\epsilon_2$ (red solid curve) during inflation. Values of the model parameters are given in~\eqref{modelparam}.
\label{Fig2slr}}
\end{figure}

For the values of models parameters
\begin{equation}
\label{modelparam}
\begin{split}
&V_0=10^{-10}M_\mathrm{Pl}^4,\quad \alpha=-0.37,\quad \beta=-1.8,\\
&\chi_0=3.5\,M_\mathrm{Pl},\quad c_0=12,\quad c_2=147,\quad d=10^{-3},
\end{split}
\end{equation}
we get inflationary parameters
 \begin{equation}
   n_s=0.9622, \quad r=0.0266, \quad A_s=2.10\cdot 10^{-9}\,,
 \end{equation}
which values are in agreement with the recent observation data~\cite{Galloni:2022mok}:
\begin{equation}
 n_s= 0.9649\pm 0.0048, \quad  r < 0.028, \quad  A_s= (2.10 \pm 0.03)\cdot 10^{-9}.
 \end{equation}

In Fig.~\ref{Fig1Hfields}, one can see that the field $\phi$ plays the role of inflaton in the beginning of inflation ($0<N<N_{*}\thickapprox 37$). After this, $\phi$ is close to zero and $\chi$ starts to increase.
So, the inflation is divided to two stages. At the first stage $N\in(0,N_{*})$, the field $\phi$ changes, tending to zero, and the field $\chi$ is almost zero. At the second stage $N\in(N_{*},N_{tot})$, the field $\phi$ is very small and the field~$\chi$ raises. The evolution of the Universe is quasi de Sitter at the beginning of both stages of inflation (see the left picture of Fig.~\ref{Fig1Hfields}).

The slow-roll parameter $\epsilon_1<1$ during inflation. In the slow-roll regime, $|\epsilon_2|<1$ as well. One can see in Fig.~\ref{Fig2slr} that the slow-roll regime is violated at the interval $37<N<41$. When the first stage of inflation ends, sufficiently large peaks appear in the power spectrum of scalar perturbations, which lead to the possible PBHs after inflation.

Following to Ref.~\cite{Geller:2022nkr}, we estimate the value of $N_{*}$ by the relation:
\begin{equation}
2\epsilon_1(N_{*})-\frac{\epsilon_2(N_{*})}{2}\simeq3
\end{equation}
We suppose that the second stage of inflation leading to the possible generation of PBH begins at the point $N_{*}$. To get the duration of inflation $N_{tot}$ we use expression $\epsilon_1(N_{tot})=1$.
\begin{table}
  \centering
      \begin{tabular}{|c|c|c|c|c|c|c|c|}
      \hline
      $\alpha$ & $\beta$ & $\phi_0/M_\mathrm{Pl}$ & $n_s$ & $r$ &
       $N_{*}$ & $N_{tot}$   \\
      \hline
       $-0.40$ & $-2$ & $5.867$ &  $0.962$ & $0.027$ & $38.2$ & $61.9$  \\
      $-0.40$ & $-1.8$ & $5.502$ &  $0.959$ & $0.027$ & $35.4$ & $58.9$\\
      $-0.40$ &  $-1.5$ & $4.936$ & $0.954$ & $0.027$ & $31.1$ & $53.5$\\
       \hline
      $-0.37$ &  $-2$ & $6.008$ & $0.965$ & $0.026$ & $40.8$ & $64.2$ \\
      $-0.37$ & $-1.8$ & $5.623$ &  $0.962$ & $0.026$ & $37.6$  & $60.7$  \\
      $-0.37$ &$-1.5$ &  $5.018$ &  $0.957$ & $0.028$ & $32.8$ & $54.9$ \\

       \hline
     $-0.35$ &$-2$ &  $6.104$ &  $0.967$ & $0.025$ & $42.7$ & $65.4$ \\
      $-0.35$ & $-1.8$ & $5.701$ &  $0.964$ & $0.026$ & $39.2$  & $61.9$  \\
      $-0.35$ & $-1.5$ & $5.072$ &  $0.959$ & $0.027$ &  $34.1$  & $56.2$  \\
      \hline
    \end{tabular}
  \caption{Dependence of inflation parameters, duration of the first stage of inflation $N_{*}$ and total duration of inflation $N_{tot}$ on the model parameters $\alpha$ and $\beta$. Other model parameters are chosen as follows: $\quad$  $V_0=10^{-10}M_\mathrm{Pl}^4,\quad\chi_0=3.5\,M_\mathrm{Pl},\quad c_0=12,\quad c_2=147,\quad d=0.003$.
  \label{TablParam}}
\end{table}

The choice of parameters of the inflation scenario under consideration specified by formula~(\ref{modelparam}) is not the only possible one. Table~\ref{TablParam} contains examples of choosing other parameter values and shows the dependence of the inflationary parameters, the duration of the first stage of inflation $N_{*}$ and the total duration of inflation $N_{tot}$ on the model parameters $\alpha$ and $\beta$. It is easy to see that the model does not contradict the observational data for $\beta=-0.2$ and $\beta=-0.18$, but for $\beta=-0.15$ and the chosen values of the parameter $\alpha$ and other parameters, the model is no longer suitable. The value of the field $\phi_0$ is chosen such that $A_s(\phi_0)=2.1\cdot 10^{-9}$.

The parameter $d$ does not influence to values of inflationary parameters, but influences to the length of inflation, namely, to the length of the second stage. There exists the suggestion \cite{Pi:2017gih} that PBH mass is related to the length of the second stage of inflation. We are interesting to estimate mass of PBHs which could be formed during radiation dominate stage. The formula for calculation of PBH mass has been proposed in Ref.~\cite{Pi:2017gih} (see also Ref.~\cite{Saburov:2023buy}):
\begin{equation}\label{25}
M_{PBH}\approx\frac{M^2_\mathrm{Pl}}{H(N_{*})}\exp\left(2(N_{tot}-N_{*})+\int\limits^{t_{tot}}_{t_{*}}\epsilon_1(t)H(t)dt\right)
\end{equation}
We substitute $\epsilon_1\,H={}-\frac{d\ln(H)}{dt}$  into Eq.~\eqref{25} and get:
\begin{equation}
M_{PBH}\approx\frac{M^2_\mathrm{Pl}}{H(N_{tot})}\exp\left(2(N_{tot}-N_{*})\right)\,.
\end{equation}

\begin{table}
  \centering
      \begin{tabular}{|c|c|c|c|c|c|}
      \hline
      $d$ & $\,N_{tot}\,$ & $N_{tot}-N_{*}$ &  $M_{PBH}/M_\mathrm{Pl}$ &$M_{PBH}/M_{\bigodot}$ & $M_{PBH}/g$ \\
      \hline
      $0.001$ &$64.5$ &  $26.9$ &  $8.57\cdot10^{28}$ & $1.87\cdot10^{-10}$ & $3.72\cdot10^{23}$ \\
      $0.002$ & $62.1$ & $24.5$ &  $7.07\cdot10^{26}$ & $1.54\cdot10^{-12}$ & $3.07\cdot10^{21}$  \\
      $0.003$ & $60.7$ & $23.1$ &  $4.31\cdot10^{25}$ & $9.40\cdot10^{-14}$ & $1.87\cdot10^{20}$ \\
      $0.007$ &  $57.7$ & $20.1$ & $1.18\cdot10^{23}$ & $2.57\cdot10^{-16}$ & $5.14\cdot10^{17}$ \\
      $0.01$ & $56.5$ & $18.9$ &  $9.92\cdot10^{21}$ & $2.16\cdot10^{-17}$ & $4.30\cdot10^{16}$ \\
      \hline
    \end{tabular}
  \caption{The dependence of duration of inflation $N_{tot}$ and the PBH mass $M_{PBH}$ from the model parameter~$d$. Other model parameters are given by (\ref{modelparam}). The end of the first stage of inflation is at $N_{*}=37.6$ independent on $d$.
  \label{Tabl}}
\end{table}

The current estimation of the mass region of PBHs considered as candidates for dark matter is $10^{-17}\,M_{\bigodot}\leqslant M_{PBH}\leqslant 10^{-12} M_{\bigodot}$, where $M_{\bigodot}$ is the Solar mass   (see Ref.~\cite{Ozsoy:2023ryl} and references therein).
As shown in Table~\ref{Tabl}, the proposed model with  $0.002\leqslant d\leqslant 0.01$ allows us to reproduce the masses of the PBH from this interval.

\section{Conclusions}

In this paper, we have proposed an  induced gravity inflationary model with two scalar fields. Using conformal transformation of the metric, we get the corresponding chiral cosmological model with two scalar fields, which assumes the formation of PBHs after inflation. The choice of parameters allows us to get the black hole masses suitable for consideration of the obtained PBHs as dark matter candidates.

Note that induced gravity term arises in inflationary models connected with particle physics~\cite{Barvinsky:1994hx,Cervantes-Cota:1995ehs,Dvali:1996ub,Bezrukov:2007ep,DeSimone:2008ei,Barvinsky:2008ia,Barvinsky:2009ii,Ferrara:2010yw,Bezrukov:2010jz,Greenwood:2012aj,Bezrukov:2013fka,Elizalde:2014xva,Elizalde:2015nya,Pozdeeva:2016hrz,Tronconi:2017wps,Dubinin:2017irg}. The form of the functions $F_1(\chi)$ and $F_2(\chi)$ is standard~\cite{Braglia:2022phb}, while the dependence of the potential on $\sigma$ is new. Note that the choice of the potential $V$ is not determined by the particle physics model, so this model can be considered a toy one. We hope that the proposed model will be useful for construction of more realistic models, unifying inflation and PBH production and motivated by particle physics.
For future investigations, it is interesting to consider processes during and after inflation  both in the Jordan frame and in the Einstein frame.

\section*{Funding}
This study was conducted within the scientific program of the National Center for Physics and Mathematics, section 5 'Particle Physics and Cosmology'. Stage 2023--2025.

%\label{sec:conflict}
%\section*{Conflict of interest}
%The authors declare no conflict of interest.

%\nocite{*}
\bibliographystyle{pepan}
\bibliography{TwoFieldInspireGravity}

\begin{thebibliography}{10}
\def\selectlanguageifdefined#1{
\expandafter\ifx\csname date#1\endcsname\relax
\else\selectlanguage{#1}\fi}
\providecommand*{\href}[2]{{\small #2}}
\providecommand*{\url}[1]{{\small #1}}
\providecommand*{\BibUrl}[1]{\url{#1}}
\providecommand{\BibAnnote}[1]{}
\providecommand*{\BibEmph}[1]{\emph{#1}}
\ProvideTextCommandDefault{\cyrdash}{\hbox to.8em{--\hss--}}
\providecommand*{\BibDash}{\ifdim\lastskip>0pt\unskip\nobreak\hskip.2em\fi
\cyrdash\hskip.2em\ignorespaces}

\bibitem{Dolgov:1992pu}
\selectlanguageifdefined{english}
\BibEmph{Dolgov A., Silk J.} {Baryon isocurvature fluctuations at small scales
  and baryonic dark matter}~//
  \href{http://dx.doi.org/10.1103/PhysRevD.47.4244}{Phys. Rev. D}. \BibDash
\newblock 1993. \BibDash
\newblock V.~47. \BibDash
\newblock P.~4244--4255.

\bibitem{Ivanov:1994pa}
\selectlanguageifdefined{english}
\BibEmph{Ivanov P., Naselsky P., Novikov I.} {Inflation and primordial black
  holes as dark matter}~//
  \href{http://dx.doi.org/10.1103/PhysRevD.50.7173}{Phys. Rev. D}. \BibDash
\newblock 1994. \BibDash
\newblock V.~50. \BibDash
\newblock P.~7173--7178.

\bibitem{Garcia-Bellido:1996mdl}
\selectlanguageifdefined{english}
\BibEmph{Garcia-Bellido J., Linde A.D., Wands D.} {Density perturbations and
  black hole formation in hybrid inflation}~//
  \href{http://dx.doi.org/10.1103/PhysRevD.54.6040}{Phys. Rev. D}. \BibDash
\newblock 1996. \BibDash
\newblock V.~54. \BibDash
\newblock P.~6040--6058. \BibDash
\newblock arXiv:astro-ph/9605094.

\bibitem{Khlopov:2008qy}
\selectlanguageifdefined{english}
\BibEmph{Khlopov M.Y.} {Primordial Black Holes}~//
  \href{http://dx.doi.org/10.1088/1674-4527/10/6/001}{Res. Astron. Astrophys.}
  \BibDash
\newblock 2010. \BibDash
\newblock V.~10. \BibDash
\newblock P.~495--528. \BibDash
\newblock arXiv:0801.0116~[astro-ph].

\bibitem{Kamenshchik:2018sig}
\selectlanguageifdefined{english}
\BibEmph{Kamenshchik A.Y., Tronconi A., Vardanyan T., Venturi G.}
  {Non-Canonical Inflation and Primordial Black Holes Production}~//
  \href{http://dx.doi.org/10.1016/j.physletb.2019.02.036}{Phys. Lett. B}.
  \BibDash
\newblock 2019. \BibDash
\newblock V. 791. \BibDash
\newblock P.~201--205. \BibDash
\newblock arXiv:1812.02547.

\bibitem{Green:2020jor}
\selectlanguageifdefined{english}
\BibEmph{Green A.M., Kavanagh B.J.} {Primordial Black Holes as a dark matter
  candidate}~// \href{http://dx.doi.org/10.1088/1361-6471/abc534}{J. Phys. G}.
  \BibDash
\newblock 2021. \BibDash
\newblock V.~48, no.~4. \BibDash
\newblock P.~043001. \BibDash
\newblock arXiv:2007.10722.

\bibitem{Carr:2020xqk}
\selectlanguageifdefined{english}
\BibEmph{Carr B., Kuhnel F.} {Primordial Black Holes as Dark Matter: Recent
  Developments}~//
  \href{http://dx.doi.org/10.1146/annurev-nucl-050520-125911}{Ann. Rev. Nucl.
  Part. Sci.} \BibDash
\newblock 2020. \BibDash
\newblock V.~70. \BibDash
\newblock P.~355--394. \BibDash
\newblock arXiv:2006.02838.

\bibitem{Dolgov:2017aec}
\selectlanguageifdefined{english}
\BibEmph{Dolgov A.D.} {Massive and supermassive black holes in the contemporary
  and early Universe and problems in cosmology and astrophysics}~//
  \href{http://dx.doi.org/10.3367/UFNe.2017.06.038153}{Usp. Fiz. Nauk}.
  \BibDash
\newblock 2018. \BibDash
\newblock V. 188, no.~2. \BibDash
\newblock P.~121--142. \BibDash
\newblock arXiv:1705.06859.

\bibitem{Ozsoy:2023ryl}
\selectlanguageifdefined{english}
\BibEmph{\"Ozsoy O., Tasinato G.} {Inflation and Primordial Black Holes}~//
  \href{http://dx.doi.org/10.3390/universe9050203}{Universe}. \BibDash
\newblock 2023. \BibDash
\newblock V.~9, no.~5. \BibDash
\newblock P.~203. \BibDash
\newblock arXiv:2301.03600.

\bibitem{Chataignier:2023ago}
\selectlanguageifdefined{english}
\BibEmph{Chataignier L., Kamenshchik A.Y., Tronconi A., Venturi G.}
  {Reconstruction methods and the amplification of the inflationary
  spectrum}~// \href{http://dx.doi.org/10.1103/PhysRevD.107.083506}{Phys. Rev.
  D}. \BibDash
\newblock 2023. \BibDash
\newblock V. 107, no.~8. \BibDash
\newblock P.~083506. \BibDash
\newblock arXiv:2301.04477.

\bibitem{Firouzjahi:2023ahg}
\selectlanguageifdefined{english}
\BibEmph{Firouzjahi H., Riotto A.} {Primordial Black Holes and loops in
  single-field inflation}~//
  \href{http://dx.doi.org/10.1088/1475-7516/2024/02/021}{JCAP}. \BibDash
\newblock 2024. \BibDash
\newblock V.~02. \BibDash
\newblock P.~021. \BibDash
\newblock arXiv:2304.07801.

\bibitem{Pi:2017gih}
\selectlanguageifdefined{english}
\BibEmph{Pi S., Zhang Y.l., Huang Q.G., Sasaki M.} {Scalaron from $R^2$-gravity
  as a heavy field}~//
  \href{http://dx.doi.org/10.1088/1475-7516/2018/05/042}{JCAP}. \BibDash
\newblock 2018. \BibDash
\newblock V.~05. \BibDash
\newblock P.~042. \BibDash
\newblock arXiv:1712.09896.

\bibitem{Gundhi:2020kzm}
\selectlanguageifdefined{english}
\BibEmph{Gundhi A., Ketov S.V., Steinwachs C.F.} {Primordial black hole dark
  matter in dilaton-extended two-field Starobinsky inflation}~//
  \href{http://dx.doi.org/10.1103/PhysRevD.103.083518}{Phys. Rev. D}. \BibDash
\newblock 2021. \BibDash
\newblock V. 103, no.~8. \BibDash
\newblock P.~083518. \BibDash
\newblock arXiv:2011.05999.

\bibitem{Braglia:2020eai}
\selectlanguageifdefined{english}
\BibEmph{Braglia M., Hazra D.K., Finelli F., Smoot G.F., Sriramkumar L.,
  Starobinsky A.A.} {Generating PBHs and small-scale GWs in two-field models of
  inflation}~// \href{http://dx.doi.org/10.1088/1475-7516/2020/08/001}{JCAP}.
  \BibDash
\newblock 2020. \BibDash
\newblock V.~08. \BibDash
\newblock P.~001. \BibDash
\newblock arXiv:2005.02895.

\bibitem{Ketov:2021fww}
\selectlanguageifdefined{english}
\BibEmph{Ketov S.V.} {Multi-Field versus Single-Field in the Supergravity
  Models of Inflation and Primordial Black Holes}~//
  \href{http://dx.doi.org/10.3390/universe7050115}{Universe}. \BibDash
\newblock 2021. \BibDash
\newblock V.~7, no.~5. \BibDash
\newblock P.~115.

\bibitem{Braglia:2022phb}
\selectlanguageifdefined{english}
\BibEmph{Braglia M., Linde A., Kallosh R., Finelli F.} {Hybrid
  \ensuremath{\alpha}-attractors, primordial black holes and gravitational wave
  backgrounds}~// \href{http://dx.doi.org/10.1088/1475-7516/2023/04/033}{JCAP}.
  \BibDash
\newblock 2023. \BibDash
\newblock V.~04. \BibDash
\newblock P.~033. \BibDash
\newblock arXiv:2211.14262.

\bibitem{Cheong:2022gfc}
\selectlanguageifdefined{english}
\BibEmph{Cheong D.Y., Kohri K., Park S.C.} {The inflaton that could: primordial
  black holes and second order gravitational waves from tachyonic instability
  induced in Higgs-R $^{2}$ inflation}~//
  \href{http://dx.doi.org/10.1088/1475-7516/2022/10/015}{JCAP}. \BibDash
\newblock 2022. \BibDash
\newblock V.~10. \BibDash
\newblock P.~015. \BibDash
\newblock arXiv:2205.14813.

\bibitem{Kawai:2022emp}
\selectlanguageifdefined{english}
\BibEmph{Kawai S., Kim J.} {Primordial black holes and gravitational waves from
  nonminimally coupled supergravity inflation}~//
  \href{http://dx.doi.org/10.1103/PhysRevD.107.043523}{Phys. Rev. D}. \BibDash
\newblock 2023. \BibDash
\newblock V. 107, no.~4. \BibDash
\newblock P.~043523. \BibDash
\newblock arXiv:2209.15343.

\bibitem{Elizalde:1993ew}
\selectlanguageifdefined{english}
\BibEmph{Elizalde E., Odintsov S.D.} {Renormalization group improved effective
  Lagrangian for interacting theories in curved space-time}~//
  \href{http://dx.doi.org/10.1016/0370-2693(94)90464-2}{Phys. Lett. B}.
  \BibDash
\newblock 1994. \BibDash
\newblock V. 321. \BibDash
\newblock P.~199--204. \BibDash
\newblock arXiv:hep-th/9311087.

\bibitem{Elizalde:2015nya}
\selectlanguageifdefined{english}
\BibEmph{Elizalde E., Odintsov S.D., Pozdeeva E.O., Vernov S.Y.} {Cosmological
  attractor inflation from the RG-improved Higgs sector of finite gauge
  theory}~// \href{http://dx.doi.org/10.1088/1475-7516/2016/02/025}{JCAP}.
  \BibDash
\newblock 2016. \BibDash
\newblock V.~02. \BibDash
\newblock P.~025. \BibDash
\newblock arXiv:1509.08817.

\bibitem{Tronconi:2017wps}
\selectlanguageifdefined{english}
\BibEmph{Tronconi A.} {Asymptotically Safe Non-Minimal Inflation}~//
  \href{http://dx.doi.org/10.1088/1475-7516/2017/07/015}{JCAP}. \BibDash
\newblock 2017. \BibDash
\newblock V.~07. \BibDash
\newblock P.~015. \BibDash
\newblock arXiv:1704.05312.

\bibitem{Kamenshchik:2019alc}
\selectlanguageifdefined{english}
\BibEmph{Kamenshchik A.Y., Tronconi A., Venturi G.} {Induced Gravity and
  Quantum Cosmology}~//
  \href{http://dx.doi.org/10.1103/PhysRevD.100.023521}{Phys. Rev. D}. \BibDash
\newblock 2019. \BibDash
\newblock V. 100, no.~2. \BibDash
\newblock P.~023521. \BibDash
\newblock arXiv:1905.02454.

\bibitem{Kamenshchik:2024kay}
\selectlanguageifdefined{english}
\BibEmph{Kamenshchik A.Y., Pozdeeva E.O., Tribolet A., Tronconi A., Venturi G.,
  Vernov S.Y.} {The Superpotential Method and the Amplification of Inflationary
  Perturbations}. \BibDash
\newblock 2024. \BibDash 6. \BibDash
\newblock arXiv:2406.19762.

\bibitem{Chervon:1995jx}
\selectlanguageifdefined{english}
\BibEmph{Chervon S.V.} {On the chiral model of cosmological inflation}~//
  \href{http://dx.doi.org/10.1007/BF00559313}{Russ. Phys. J.} \BibDash
\newblock 1995. \BibDash
\newblock V.~38. \BibDash
\newblock P.~539--543.

\bibitem{Karananas:2016kyt}
\selectlanguageifdefined{english}
\BibEmph{Karananas G.K., Rubio J.} {On the geometrical interpretation of
  scale-invariant models of inflation}~//
  \href{http://dx.doi.org/10.1016/j.physletb.2016.08.037}{Phys. Lett. B}.
  \BibDash
\newblock 2016. \BibDash
\newblock V. 761. \BibDash
\newblock P.~223--228. \BibDash
\newblock arXiv:1606.08848.

\bibitem{Chervon:2019nwq}
\selectlanguageifdefined{english}
\BibEmph{Chervon S.V., Fomin I.V., Pozdeeva E.O., Sami M., Vernov S.Y.}
  {Superpotential method for chiral cosmological models connected with modified
  gravity}~// \href{http://dx.doi.org/10.1103/PhysRevD.100.063522}{Phys. Rev.
  D}. \BibDash
\newblock 2019. \BibDash
\newblock V. 100, no.~6. \BibDash
\newblock P.~063522. \BibDash
\newblock arXiv:1904.11264.

\bibitem{Fomin:2020caa}
\selectlanguageifdefined{english}
\BibEmph{Fomin I., Chervon S.} {Exact and Slow-Roll Solutions for Exponential
  Power-Law Inflation Connected with Modified Gravity and Observational
  Constraints}~// \href{http://dx.doi.org/10.3390/universe6110199}{Universe}.
  \BibDash
\newblock 2020. \BibDash
\newblock V.~6, no.~11. \BibDash
\newblock P.~199.

\bibitem{Geller:2022nkr}
\selectlanguageifdefined{english}
\BibEmph{Geller S.R., Qin W., McDonough E., Kaiser D.I.} {Primordial black
  holes from multifield inflation with nonminimal couplings}~//
  \href{http://dx.doi.org/10.1103/PhysRevD.106.063535}{Phys. Rev. D}. \BibDash
\newblock 2022. \BibDash
\newblock V. 106, no.~6. \BibDash
\newblock P.~063535. \BibDash
\newblock arXiv:2205.04471.

\bibitem{Pozdeeva:2024lah}
\selectlanguageifdefined{english}
\BibEmph{Pozdeeva E.O., Vernov S.Y.} {Construction of Chiral Cosmological
  Models Unifying Inflation and Primordial Black Hole Formation}~//
  \href{http://dx.doi.org/10.17238/issn2226-8812.2024.1.90-94}{SPACE, TIME AND
  FUNDAMENTAL INTERACTIONS}. \BibDash
\newblock 2024. \BibDash 1. \BibDash
\newblock V. 1(46). \BibDash
\newblock P.~90--94. \BibDash
\newblock arXiv:2401.12040.

\bibitem{Galloni:2022mok}
\selectlanguageifdefined{english}
\BibEmph{Galloni G., Bartolo N., Matarrese S., Migliaccio M., Ricciardone A.,
  Vittorio N.} {Updated constraints on amplitude and tilt of the tensor
  primordial spectrum}~//
  \href{http://dx.doi.org/10.1088/1475-7516/2023/04/062}{JCAP}. \BibDash
\newblock 2023. \BibDash
\newblock V.~04. \BibDash
\newblock P.~062. \BibDash
\newblock arXiv:2208.00188.

\bibitem{Saburov:2023buy}
\selectlanguageifdefined{english}
\BibEmph{Saburov S., Ketov S.V.} {Improved Model of Primordial Black Hole
  Formation after Starobinsky Inflation}~//
  \href{http://dx.doi.org/10.3390/universe9070323}{Universe}. \BibDash
\newblock 2023. \BibDash
\newblock V.~9, no.~7. \BibDash
\newblock P.~323. \BibDash
\newblock arXiv:2306.06597.

\bibitem{Barvinsky:1994hx}
\selectlanguageifdefined{english}
\BibEmph{Barvinsky A.O., Kamenshchik A.Y.} {Quantum scale of inflation and
  particle physics of the early universe}~//
  \href{http://dx.doi.org/10.1016/0370-2693(94)91253-X}{Phys. Lett. B}.
  \BibDash
\newblock 1994. \BibDash
\newblock V. 332. \BibDash
\newblock P.~270--276. \BibDash
\newblock arXiv:gr-qc/9404062.

\bibitem{Cervantes-Cota:1995ehs}
\selectlanguageifdefined{english}
\BibEmph{Cervantes-Cota J.L., Dehnen H.} {Induced gravity inflation in the
  standard model of particle physics}~//
  \href{http://dx.doi.org/10.1016/0550-3213(95)00128-X}{Nucl. Phys. B}.
  \BibDash
\newblock 1995. \BibDash
\newblock V. 442. \BibDash
\newblock P.~391--412. \BibDash
\newblock arXiv:astro-ph/9505069.

\bibitem{Dvali:1996ub}
\selectlanguageifdefined{english}
\BibEmph{Dvali G.R.} {Natural inflation in SUSY and gauge mediated curvature of
  the flat directions}~//
  \href{http://dx.doi.org/10.1016/0370-2693(96)01063-5}{Phys. Lett. B}.
  \BibDash
\newblock 1996. \BibDash
\newblock V. 387. \BibDash
\newblock P.~471--477. \BibDash
\newblock arXiv:hep-ph/9605445.

\bibitem{Bezrukov:2007ep}
\selectlanguageifdefined{english}
\BibEmph{Bezrukov F.L., Shaposhnikov M.} {The Standard Model Higgs boson as the
  inflaton}~// \href{http://dx.doi.org/10.1016/j.physletb.2007.11.072}{Phys.
  Lett. B}. \BibDash
\newblock 2008. \BibDash
\newblock V. 659. \BibDash
\newblock P.~703--706. \BibDash
\newblock arXiv:0710.3755~[hep-th].

\bibitem{DeSimone:2008ei}
\selectlanguageifdefined{english}
\BibEmph{De~Simone A., Hertzberg M.P., Wilczek F.} {Running Inflation in the
  Standard Model}~//
  \href{http://dx.doi.org/10.1016/j.physletb.2009.05.054}{Phys. Lett. B}.
  \BibDash
\newblock 2009. \BibDash
\newblock V. 678. \BibDash
\newblock P.~1--8. \BibDash
\newblock arXiv:0812.4946~[hep-ph].

\bibitem{Barvinsky:2008ia}
\selectlanguageifdefined{english}
\BibEmph{Barvinsky A.O., Kamenshchik A.Y., Starobinsky A.A.} {Inflation
  scenario via the Standard Model Higgs boson and LHC}~//
  \href{http://dx.doi.org/10.1088/1475-7516/2008/11/021}{JCAP}. \BibDash
\newblock 2008. \BibDash
\newblock V.~11. \BibDash
\newblock P.~021. \BibDash
\newblock arXiv:0809.2104~[hep-ph].

\bibitem{Barvinsky:2009ii}
\selectlanguageifdefined{english}
\BibEmph{Barvinsky A.O., Kamenshchik A.Y., Kiefer C., Starobinsky A.A.,
  Steinwachs C.F.} {Higgs boson, renormalization group, and naturalness in
  cosmology}~// \href{http://dx.doi.org/10.1140/epjc/s10052-012-2219-3}{Eur.
  Phys. J. C}. \BibDash
\newblock 2012. \BibDash
\newblock V.~72. \BibDash
\newblock P.~2219. \BibDash
\newblock arXiv:0910.1041~[hep-ph].

\bibitem{Ferrara:2010yw}
\selectlanguageifdefined{english}
\BibEmph{Ferrara S., Kallosh R., Linde A., Marrani A., Van~Proeyen A.} {Jordan
  Frame Supergravity and Inflation in NMSSM}~//
  \href{http://dx.doi.org/10.1103/PhysRevD.82.045003}{Phys. Rev. D}. \BibDash
\newblock 2010. \BibDash
\newblock V.~82. \BibDash
\newblock P.~045003. \BibDash
\newblock arXiv:1004.0712~[hep-th].

\bibitem{Bezrukov:2010jz}
\selectlanguageifdefined{english}
\BibEmph{Bezrukov F., Magnin A., Shaposhnikov M., Sibiryakov S.} {Higgs
  inflation: consistency and generalisations}~//
  \href{http://dx.doi.org/10.1007/JHEP01(2011)016}{JHEP}. \BibDash
\newblock 2011. \BibDash
\newblock V.~01. \BibDash
\newblock P.~016. \BibDash
\newblock arXiv:1008.5157~[hep-ph].

\bibitem{Greenwood:2012aj}
\selectlanguageifdefined{english}
\BibEmph{Greenwood R.N., Kaiser D.I., Sfakianakis E.I.} {Multifield Dynamics of
  Higgs Inflation}~// \href{http://dx.doi.org/10.1103/PhysRevD.87.064021}{Phys.
  Rev. D}. \BibDash
\newblock 2013. \BibDash
\newblock V.~87. \BibDash
\newblock P.~064021. \BibDash
\newblock arXiv:1210.8190~[hep-ph].

\bibitem{Bezrukov:2013fka}
\selectlanguageifdefined{english}
\BibEmph{Bezrukov F.} {The Higgs field as an inflaton}~//
  \href{http://dx.doi.org/10.1088/0264-9381/30/21/214001}{Class. Quant. Grav.}
  \BibDash
\newblock 2013. \BibDash
\newblock V.~30. \BibDash
\newblock P.~214001. \BibDash
\newblock arXiv:1307.0708~[hep-ph].

\bibitem{Elizalde:2014xva}
\selectlanguageifdefined{english}
\BibEmph{Elizalde E., Odintsov S.D., Pozdeeva E.O., Vernov S.Y.}
  {Renormalization-group improved inflationary scalar electrodynamics and SU(5)
  scenarios confronted with Planck 2013 and BICEP2 results}~//
  \href{http://dx.doi.org/10.1103/PhysRevD.90.084001}{Phys. Rev. D}. \BibDash
\newblock 2014. \BibDash
\newblock V.~90, no.~8. \BibDash
\newblock P.~084001. \BibDash
\newblock arXiv:1408.1285~[hep-th].

\bibitem{Pozdeeva:2016hrz}
\selectlanguageifdefined{english}
\BibEmph{Pozdeeva E.O., Vernov S.Y.} {Renormalization-group improved
  inflationary scenarios}~//
  \href{http://dx.doi.org/10.1134/S1547477117020273}{Phys. Part. Nucl. Lett.}
  \BibDash
\newblock 2017. \BibDash
\newblock V.~14, no.~2. \BibDash
\newblock P.~386--389. \BibDash
\newblock arXiv:1604.02272.

\bibitem{Dubinin:2017irg}
\selectlanguageifdefined{english}
\BibEmph{Dubinin M.N., Petrova E.Y., Pozdeeva E.O., Sumin M.V., Vernov S.Y.}
  {MSSM-inspired multifield inflation}~//
  \href{http://dx.doi.org/10.1007/JHEP12(2017)036}{JHEP}. \BibDash
\newblock 2017. \BibDash
\newblock V.~12. \BibDash
\newblock P.~036. \BibDash
\newblock arXiv:1705.09624.

\end{thebibliography}

\end{document}